\documentclass[11pt]{article}
\usepackage{cite}
\usepackage{scalerel}
\usepackage{xcolor}
\usepackage{shuffle}
\usepackage{amsmath,amsfonts,amssymb}
\usepackage{latexsym,epsfig}
\usepackage{dsfont}
\usepackage{hyperref}
\usepackage{extarrows}
\usepackage{comment} 
\usepackage{tikz}
\usepackage{faktor}
\usetikzlibrary{decorations.pathmorphing}
\usepackage{mathtools}
\tikzset{snake it/.style={decorate, decoration=snake}}

\newcommand{\mycomment}[1]{}

\def\hybrid{
        \topmargin -20pt
        \oddsidemargin 0pt
        \headheight 0pt \headsep 0pt
        \textwidth 6.25in 
        \textheight 9.5in 
        \marginparwidth .875in
        \parskip 5pt plus 1pt \jot = 1.5ex}

\hybrid

 \csname
@addtoreset\endcsname{equation}{section}

\newcommand{\be}{\begin{equation}}
\newcommand{\ee}{\end{equation}}

\newcommand{\ba}{\begin{equation} \begin{aligned}}
\newcommand{\ea}{\end{aligned} \end{equation}}

\def\cC{{\cal C}}
\def\cB{{\cal B}}
\def\cG{{\cal G}}
\def\cJ{{\cal J}}

\def\cF{{\cal F}}
\def\cO{{\cal O}}
\def\cA{{\cal A}}
\def\cE{{\cal E}}
\def\cM{{\cal M}}
\def\cN{{\cal N}}
\def\cQ{{\cal Q}}
\def\cR{{\cal R}}

\def\cP{{\cal P}}

\def\cK{{\cal K}}

\def\cH{{\cal H}}

\def\cW{{\cal W}}
\def\del{\partial}

\def\B{\square}

\allowdisplaybreaks

\def\bpm{\begin{pmatrix}}
\def\epm{\end{pmatrix}}

\begin{document}

\begin{titlepage}
\rightline{}
\rightline{August  2025}
\rightline{HU-EP-25/30-RTG}  
\begin{center}
\vskip 1.5cm
{\Large \bf{Yang-Mills kinematic algebra via homotopy transfer \\[1ex]from a worldline operator algebra}}

\vskip 1.7cm

{\large\bf {Roberto Bonezzi, Christoph Chiaffrino, Olaf Hohm, Maria Foteini Kallimani}}
\vskip 1.6cm

{\it  Institute for Physics, Humboldt University Berlin,\\
 Zum Gro\ss en Windkanal 2, D-12489 Berlin, Germany}\\[1.5ex] 
 roberto.bonezzi@physik.hu-berlin.de, chiaffrc@hu-berlin.de,\\ ohohm@physik.hu-berlin.de, kallimari@physik.hu-berlin.de
\vskip .1cm

\vskip .2cm

\end{center}

\bigskip\bigskip
\begin{center} 
\textbf{Abstract}

\end{center} 
\begin{quote}

The homotopy Lie or $L_{\infty}$ algebra encoding Yang-Mills theory is the tensor product of a color Lie algebra with the kinematic $C_{\infty}$ algebra.  
We derive this $C_{\infty}$ algebra, via homotopy transfer,  from a strict  operator algebra of a worldline theory, realized as an associative star product algebra. This gives a homotopy transfer interpretation to worldline vertex operators introduced in previous work.

\end{quote} 
\vfill
\setcounter{footnote}{0}
\end{titlepage}

\tableofcontents

\section{Introduction}

The kinematic algebra of Yang-Mills theory denotes a conjectural structure  suggested by the color-kinematics duality discovered at the level of scattering amplitudes, which in turn is the basis for the double copy construction of gravity amplitudes as `the square' of gauge theory amplitudes \cite{Bern:2008qj,Bern:2010ue,Bern:2022wqg}. While these properties are well understood at the level of on-shell objects like tree-level scattering amplitudes, it was only more recently that significant progress was made in finding a first-principle derivation that is off-shell, local and gauge invariant \cite{Anastasiou:2018rdx,Borsten:2020zgj,Diaz-Jaramillo:2021wtl,Bonezzi:2022yuh,Borsten:2022vtg,Bonezzi:2022bse,Bonezzi:2023pox,Borsten:2023ned,Bonezzi:2023lkx,Borsten:2023paw,Bonezzi:2024dlv}. 
In this, the formulation of field theory in terms of homotopy algebras is instrumental \cite{Zwiebach:1992ie,Lada:1992wc,Hohm:2017pnh,Borsten:2021hua}.
In particular, an algebraic structure capturing color-kinematics duality was proposed by Reiterer \cite{Reiterer:2019dys} and, being a homotopy generalization of a Batalin-Vilkovisky (BV) algebra,  was named BV$_{\infty}^{\square}$, with $\square$ referring to the d'Alembert wave operator.
Except for special theories and  formulations \cite{Reiterer:2019dys,Bonezzi:2024dlv,Borsten:2022vtg}, 
this BV$_{\infty}^{\square}$ algebra is only partially understood.

Our goal in this paper is to 
make progress in finding a general all-order description of this elusive algebra. 
One would like to derive the BV$_{\infty}^{\square}$ algebra,  including its infinitely many maps of ever increasing arity, 
from something simple, ideally an algebra that is `strict' in that all maps or brackets are at most bilinear. 
Here we will follow an approach that seems particularly promising to us, based on a worldline QFT and its operator algebra. In fact, recently three of us showed that part of the 
Yang-Mills kinematic algebra  
can be derived from the strict algebra of vertex operators associated to this worldline QFT \cite{Bonezzi:2024fhd}. 
(In this we were greatly inspired by the work of Lian-Zuckermann and Zeitlin \cite{Lian:1992mn,Zeitlin:2007vd,Zeitlin:2014xma,Zeitlin:2008cc,Zeitlin:2007vv,Zeitlin:2009tj}.)
However, these vertex operators were unconventional, requiring non-linear contributions, see also \cite{Bonezzi:2025iza,Bastianelli:2025xne} for the role of non-linear vertex operators in worldline theories. 

In this paper we will show that the vertex operators of \cite{Bonezzi:2024fhd}, notably their non-linear contributions, have an interpretation in terms of homotopy transfer from a strictly associative operator algebra. (See \cite{Crainic:2004bxw,Erbin:2020eyc,Koyama:2020qfb,Arvanitakis:2020rrk,Arvanitakis:2021ecw,Chiaffrino:2020akd,Chiaffrino:2023wxk} for introductions and applications of homotopy transfer.) 
In this the vertex operators will be interpreted as the inclusion map that injects the non-strict color-stripped Yang-Mills homotopy  algebra (a $C_{\infty}$ algebra that is a subsector of the BV$_{\infty}^{\square}$ algebra) into the strict operator algebra. 
Concretely, we will define 
 a strictly associative algebra ${\cal F}$ based on a Moyal-Weyl-type star product, using  the methods of deformation quantization  \cite{Fedosov:1994zz,Bayen:1977hb,Bayen:1977pr,Kontsevich:1997vb,Gerstenhaber:1963zz}, 
and then show that there is a homotopy transfer to  the kinematic space ${\cal  K}$.   
 Apart from the vertex operators  being  interpreted as the (non-trivial) inclusion map ${V} = I : {\cal K}\rightarrow {\cal F}$,  we have the projector $\Pi : {\cal F}\rightarrow {\cal K}$ and the homotopy $H:{\cal F}\rightarrow {\cal F}$. The star algebra is 
 transported to ${\cal K}$ via: 
  \be
   \mu_2(\psi_1, \psi_2) = \Pi\big({V}(\psi_1)\star {V}(\psi_2)\big) \;,  
  \ee
for $\psi_1, \psi_2\in {\cal K}$, which yields (up to homotopy) the kinematic $C_{\infty}$ product of Yang-Mills theory.   
Acting with ${V}$ on this equation and using the homotopy relation ${V}\Pi=I \Pi=1_{\cal F}+[{\cal Q}, H]$
one obtains 
 \be
  { V}\big(\mu_2(\psi_1, \psi_2)\big) = {V}(\psi_1)\star { V}(\psi_2) + [{\cal Q}, H]( { V}(\psi_1)\star { V}(\psi_2))\;, 
 \ee
which gives a \textit{derived} construction of the bilinear vertex operator $ { V}_{2}(\psi_1, \psi_2) = H( { V}(\psi_1)\star { V}(\psi_2)) $ previously determined by hand in \cite{Bonezzi:2024fhd}.

Let us emphasize from the outset that the central goal here is {not} so much to derive the $C_{\infty}$ algebra on the kinematic space ${\cal K}$ of Yang-Mills theory from something strict. This could simply be achieved  by making Yang-Mills theory cubic, say by introducing an auxiliary two-form field. Rather, the goal is to do so in the context of the worldline theory, as this has proven to provide the right maps suggested by string theory, including the $b$-ghost to be discussed below, which  are instrumental for the understanding of the full BV$_{\infty}^{\square}$ algebra. 
So far the derived construction is restricted to the $C_{\infty}$ subsector, but we will comment in the conclusions on possible strategies to eventually capture the full kinematic algebra. 

The rest of this paper is organized as follows. In sec.~2 we introduce the worldline theory and define its operator algebra, which we realize as a star product algebra of functions. 
Moreover, we define a state space on which this algebra acts. 
In sec.~3 we consider subspaces and ideals of this operator algebra, which is needed to define the subspace onto which the homotopy transfer maps. 
This homotopy transfer is spelled out in sec.~4. We close with an outlook in sec.~5.

\section{Worldline inspired phase space}

In this section, we review the phase space of the bosonic spinning particle, which has been treated in more detail in \cite{Bonezzi:2024emt}. We present the phase space of the gauge-fixed classical theory, which is enriched with fermionic coordinates corresponding to BRST ghosts. Quantization is performed through the deformation of the pointwise product on phase space into a star product, yielding an associative and non-commutative algebra on the quantized phase space.

\subsection{Classical phase space}

The free bosonic spinning particle is described by the following symplectic structure
\begin{equation}
    S_{\rm symp}=\int d\tau\Big[p_\mu\dot{x}^\mu+\bar{\alpha}_\mu\dot{\alpha}^\mu\Big]\,,
\end{equation}
in the Hamiltonian action. The phase space of the theory contains, first of all, the bosonic canonical pairs of the position and momentum of the particle, $(x^\mu,p_\nu)$, as well as the worldline oscillators $(\alpha^\mu,\bar{\alpha}_\nu)$ which generate spin degrees of freedom. We denote spacetime indices with Greek letters, which take values $\mu=0,\dots, D-1$. The following phase space functions
\begin{align}
H:=p^2\,,&&L:=\alpha\cdot p\,, &&\bar{L}:=\bar{\alpha}\cdot p\;,
\end{align}
form a first-class algebra of Hamiltonian constraints under Poisson brackets.
In the quantized theory, the operators associated to these constraints form a contraction of the $sl(2,\mathbb{R})$ subalgebra of the Virasoro algebra, as discussed in \cite{Bonezzi:2024emt}, and more extensively in \cite{Bouatta:2004kk,Bengtsson:1986ys}. To ensure unitarity of the quantum theory, we need to gauge these constraints, yielding the action of the bosonic spinning particle
\begin{equation}
    \label{gauged action}
    S=\int d\tau\left[p_\mu \dot{x}^\mu+\bar{\alpha}^\mu\dot{\alpha}_\mu-eH-u\bar{L}-\bar{u}L\right]\,,
\end{equation}
where $e$, $u$, $\bar{u}$ are Lagrange multipliers. For the purpose of Hamiltonian BRST quantization, we assign to each constraint a canonical pair of fermionic ghosts $(c^i, b_i)$, with $i\in\{1,2,3\}$ where $c^i:=(\cC,c,\bar{\cC})$ and $b_i:=(\bar{\cB},b,\cB)$. Upon gauge fixing $e=1$, $u=0$, $\bar{u}=0$, the full action in first-order form is given by 
\begin{equation}
    \label{full action}
    S=\int d\tau \left[p_\mu\dot{x}^\mu+\bar{\alpha}^\mu\dot{\alpha}_\mu-b_i\dot{c}^i-p^2\right]\,.
\end{equation}

All in all, we identify our graded phase space to be $\cP:=\mathbb{R}^{4D|6}$, meaning that we have $4D$ bosonic coordinates, with $D$ being the dimension of spacetime, and 6 fermionic ones. We describe this space with coordinates $(X^A,P_A)$, where
\begin{align}
    \label{coordinates}
    X^A:=(x^\mu, \alpha^\nu, c,\cC,\cB)\,,&& P_A:=(p_\mu,\bar{\alpha}_\nu, b,\bar{\cB},\bar{\cC})\,.
\end{align}
Capital Latin indices will from now on run over the phase space coordinates. Note that $P_A=0$ is a Lagrangian submanifold of the phase space $\cP$, which we will use in order to define our state space. However, any such choice will lead to an equivalent quantization of the theory. 

The phase space $\cP$ naturally admits a $\mathbb{Z}_2$ grading, induced by the fermionic or bosonic nature of each coordinate. In fact, the theory possesses a $U(1)_\cG\times U(1)_\cJ$ symmetry. The $U(1)_\cG$ sector comes from further refining our grading by assigning \textit{ghost numbers} $\cG$ as $\cG(c^i)=+1$, $\cG(b_i)=-1$, while choosing all bosonic coordinates to have degree zero. We denote indices of the ghost or antighost multiplets with lowercase Latin letters $i$ and $j$. 

Furthermore, the $U(1)_\cJ$ symmetry is related to another charge $\cJ$, which is such that $\cJ(x^\mu,p_\mu,c,b)=0$, while $\cJ(\alpha^\mu,\cC,\cB)=+1$ and $\cJ(\bar{\alpha}^\mu,\bar{\cC},\bar{\cB})=-1$. The ghost and antighost triplets, $c^i$ and $b_i$, are thus organized with decreasing and increasing charge $\cJ$, respectively. We can also reorganize our coordinates as $\mathfrak{a}^m:=(\cB, \alpha^\mu, \cC)$ with $\cJ=1$, $\bar{\mathfrak{a}}_m:=(\bar{\cC}, \bar{\alpha}_\mu, \bar{\cB})$ with $\cJ=-1$. We also have the $U(1)_\cJ$ singlets $x^\mu$, $p_\mu$, $c$ and $b$. We represent indices of the charge multiplets with lowercase Latin letters starting with $m$. The $U(1)_\cJ$ symmetry expresses  the invariance of the action \eqref{full action} under simultaneously transforming $\mathfrak{a}^m\to e^{i\theta}\mathfrak{a}^m$ and $\bar{\mathfrak{a}}_m\to e^{-i\theta}\mathfrak{a}_m$.

We restrict the class of functions on phase space to the subspace 
\begin{equation}
\label{classic phase space}
    \cO:=C^\infty(\mathbb{R}^{{D}})[p^\mu,\mathfrak{a}^m,\bar{\mathfrak{a}}_n,c,b]\,,
\end{equation}
of polynomials in $\{p^\mu,\mathfrak{a}^m,\bar{\mathfrak{a}}_n,c,b\}$ with coefficients that are smooth functions of $x^\mu$. This space is naturally equipped with the usual pointwise product of functions, which defines the graded commutative algebra $(\cO,\cdot)$. Simultaneously, it can be imbued with a Poisson bracket, defined as 
\begin{equation}
    \label{Poisson bracket}
    \left\{F,G\right\}:=(-1)^{\cG(X)}F\frac{\overleftarrow{\partial }}{\partial X^A}\frac{\overrightarrow{\partial }}{\partial P_A}G-F\frac{\overleftarrow{\partial }}{\partial P_A}\frac{\overrightarrow{\partial }}{\partial X^A}G\,,
\end{equation}
where the right- and left-derivatives are related by $F\frac{\overleftarrow{\partial}}{\partial X^A}=(-1)^{X(F+1)}\frac{\partial F}{\partial X^A}$, and from now on we will use the short-hand notation $(-1)^{\cG(F)}\equiv(-1)^F$. With these conventions, the canonical Poisson bracket is $\{P_A,X^B\}=-\delta_A^B$, which is the inverse of the symplectic structure given through the action \eqref{full action}. The algebra $(\cO,\{-,-\})$ is  a graded Poisson algebra.

The gauge fixed theory features a BRST symmetry, with an associated BRST charge of ghost number $+1$. This charge is given as
\begin{equation}
    \label{Q}
    Q:=c\,p^2+\left(\bar{\cC}\alpha^\mu+\cC\bar{\alpha}^\mu \right)p_\mu-\cC\bar{\cC} b\,.
\end{equation}
We can use this charge to construct a nilpotent differential as the Hamiltonian vector field $\{Q,-\}$, given that $\{Q,Q\}=0$.
This is a degree $+1$ differential that turns $(\cO,\{-,-\},\{Q,-\})$ into a differential graded Poisson algebra.  In particular, $(\cO,\{Q,-\})$ is a chain complex and its cohomology encodes the classical observables of the theory.

Note that, at this point, we can assign mass dimension to the phase space variables. Demanding that the BRST charge is dimensionless, $[Q]=0$, and knowing that $[p_\mu]=+1$, $[x^\mu]=-1$, as well as the requirement of antighosts to have the opposite mass dimension than the corresponding ghosts, we get the following assignments:\footnote{The only constraint for the mass dimension for $\alpha^\mu$, $\bar{\alpha}^\mu$, $\cC$, $\bar{\cC}$ is that they should satisfy $[\cC]+[\bar{\alpha}^\mu]=[\bar{\cC}]+[\alpha^\mu]=-1$, from demanding $[Q]=0$, as well as that $[\alpha^\mu]+[\bar{\alpha}^\mu]=0$ from $[S]=0$. There are other choices for the specific mass dimensions of these variables which are consistent, but they are all equivalent.}
\begin{align}
    &[c]=-2\,,&&[b]=+2\,,\\
    &[\cC]=[\bar{\cC}]=+1&&[\cB]=[\bar{\cB}]=-1\,,\\
    &[\alpha^\mu]=[\bar{\alpha}^\mu]=0\,.
\end{align}

\subsection{Deformation quantization}

To quantize our theory, we deform the graded commutative algebra $(\cO,\cdot)$ into a non-commutative graded algebra $(\cO,\star)$. The new algebra is equipped with a star product, a bilinear map
\begin{align}
    \star:\cO\times \cO \to \cO, && (F,G)\mapsto F\star  G:=\sum_{n\geq 0}\hbar^n L_n(F,G)\,,
\end{align}
where we take $\hbar$ to be a general complex number. This product has the following properties:
\begin{itemize}
    \item It is associative: $(F\star G)\star H=F\star (G\star H)$ for $F,\,G,\,H\in \cO$,
    \item There is a unit element, $1\star F=F\star 1=F$,
    \item The map $L_0$ is the pointwise product, $L_0(F,G)=F\cdot G$,
    \item The map $L_1$ is related to the Poisson bracket: $L_1(F,G)-(-1)^{FG}L_1(G, F)=i\{F,G\}$.
\end{itemize}
As the algebra $(\cO,\star)$ is associative, one can further define a graded Lie bracket by introducing the graded star commutator 
\begin{equation}
[F,G]_\star:=F\star G -(-1)^{FG} G \star F\;.    
\end{equation}
One recovers the classical algebra by taking the classical limit, $\hbar \to 0$: 
\begin{align}
    \lim_{\hbar\to 0}F\star G=F\cdot G\,, && \lim_{\hbar\to 0}\,\frac{1}{i\hbar}[F,G]_\star=\{F,G\}\,.
\end{align}
We can use the star commutator in order to define a (left) adjoint action of $(\cO,\star)$ on itself. Given $W,F\in \cO$, we define
\begin{equation}
\cW\,:\,\cO\,\longrightarrow\,\cO\;,\quad \cW(F):=[W,F]_\star\;,    
\end{equation}
so that $\cW\in{\rm End}(\cO)$ is a derivation of the star product.

The choice of star product for the algebra of functions in deformation quantization is equivalent to a choice of ordering for operators in the canonical formalism. We will elaborate on the relation of the two quantization schemes further below.
We first exemplify the relevant concepts for the simplest case of a two-dimensional bosonic phase space, with coordinates $(X,P)$, and then give their concrete realizations for our model of interest.
Two typical choices for the star product are the Weyl-Moyal product $\star_{\rm w}$ and the normal ordered product $\star_{\rm n}$, given by 
\begin{equation}\label{Weyl and Normal star product}
\begin{split}
F\star_{\rm w} G&:=F \exp\left[-\frac{i\hbar}{2}\left(\frac{\overleftarrow{\del}}{\del P} \frac{\overrightarrow{\del}}{\del X}-\frac{\overleftarrow{\del}}{\del X} \frac{\overrightarrow{\del}}{\del P}\right)\right]G\;,\\
F\star_{\rm n} G&:=F \exp\left[-i\hbar\,\frac{\overleftarrow{\del}}{\del P} \frac{\overrightarrow{\del}}{\del X}\right]G\;,
\end{split}
\end{equation}
where $F(X,P)$ and $G(X,P)$ are functions on phase space. Notice that, when acting on functions that are polynomials in $P$, the above exponentials truncate at some finite order and the star product closes on polynomials of finite, but arbitrary, order.  This ensures that the space $\cO$ is closed under the star product. With the above choices, the identity element is the constant function $1$.

The product $\star_{\rm w}$ treats $X$ and $P$ on equal footing and is related to the Weyl (symmetric) ordering of position and momentum operators in canonical quantization. The product $\star_{\rm n}$, in which we take only $X$-derivatives of $G$ and $P$-derivatives of $F$, is related to the normal ordering at the level of operators, where all momenta are moved  to the right. The two star products provide isomorphic realizations of the algebra of operators. They can be mapped to one another via
\begin{equation}
    \label{iso star w star normal}
    e^G\big(F \star_{\rm w} H\big)=(e^{G}F) \star_{\rm n} (e^{G}H)\;,\quad G=-\frac{i\hbar}{2}\,\del_X\del_P\;,
\end{equation}
where $e^G$ essentially takes Wick contractions. This is most commonly seen in quantum field theory, where the equivalent of \eqref{iso star w star normal} re-expresses time-ordered products of operators into normal-ordered ones, via Wick's theorem.

We can now define the states of the theory to be functions on a Lagrangian submanifold of the phase space. With the polarization $(X,P)$ of phase space coordinates, we can for example choose states to be given by functions of $X$.
The space of states $\cH$ then carries a representation of the star algebra of functions: given a function $F(X,P)$ on phase space, we can construct an operator $\rho(F):\cH\rightarrow\cH$, whose action on a state $\phi(X)$ we denote as $\rho(F;\phi)\in\cH$. The requirement that $\rho$ furnishes a representation of the star algebra amounts to
\begin{equation}\label{representation rho}
\rho(F)\circ\rho(G)=\rho(F\star G)\;.    
\end{equation}
Depending on the choice of star product, the representation map is different. In normal ordering one simply has
\begin{equation}\label{rho normal}
\rho_{\rm n}(F;\phi):=\big(F\star_{\rm n}\phi\big)\big\rvert_{P=0}\;,
\end{equation}
which obeys \eqref{representation rho} because $\star_{\rm n}$ does not take $P$-derivatives of the right argument. 
To use the Weyl-Moyal product, instead, one has to employ the isomorphism \eqref{iso star w star normal} to define the representation
\begin{equation}\label{rho Weyl}
\rho_{\rm w}(F;\phi):=\Big[e^G\big(F\star_{\rm w}\phi\big)\Big]\Big\rvert_{P=0}\;.    
\end{equation}

Going back to our actual model, for the purposes of this paper we will use a star product that implements a mix of Weyl and normal ordering. Specifically, we use $\star_{\rm w}$ for the $\cJ=0$ variables, while we employ $\star_{\rm n}$ for the $\cJ=\pm 1 $ variables. To that end, we introduce the product
\begin{equation}\label{our star}
F\star G:=F\exp\left[ -\frac{i\hbar}{2}\left(\overleftarrow{\partial_{p}}\cdot\overrightarrow{\partial_{x}}-\overleftarrow{\partial_{x}}\cdot\overrightarrow{\partial_{p}} +\overleftarrow{\partial_{b}}\,\overrightarrow{\partial_{c}}+\overleftarrow{\partial_{c}}\,\overrightarrow{\partial_{b}}   \right)-i\hbar\,\overleftarrow{\partial_{\bar{\mathfrak{a}}}}\cdot\overrightarrow{\partial_\mathfrak{a}}\right] G\;,    
\end{equation}
where we recall that $F$ and $G$ are functions of the graded coordinates $X^A=(x^\mu,c,\mathfrak{a}^m)$ and momenta $P_A=(p_\mu,b,\bar{\mathfrak{a}}_m)$. The star product above is the one we refer to when using the symbol $\star$ from now on. We define the graded associative algebra $\cO_q:=(\cO,\star)$, that is the space of functions \eqref{classic phase space} endowed with the star product \eqref{our star}. 

For the corresponding state space $\cH$, we choose functions $\phi(X)$ which are smooth functions of $x^\mu$ and polynomials in $\mathfrak{a}^m$ and $c$, i.e.
\begin{equation}
\cH:=C^\infty(\mathbb{R}^{{D}})[\mathfrak{a}^m,c] \;,   
\end{equation}
which is indeed the subspace of $\cO$ that does not depend on the momenta $P_A$. Given the mixed form of the star product \eqref{our star}, the action of the star algebra $\cO_q$ on $\cH$ is given by
\begin{equation}
\label{action on states}
    \rho(F;\phi):=\Big[e^{G}(F\star \phi)\Big]\Big\rvert_{P_A=0}\;,
\end{equation}
where we now extend $G$ to
\begin{equation}
    \label{Wick contractor}
    G:=-\frac{i\hbar}{2}\big(\partial_{x}\cdot \partial_{p}+\partial_c\,\partial_b\big)\;,
\end{equation}
in agreement with the Weyl representation \eqref{rho Weyl} for the $\cJ=0$ variables. Because of the algebra isomorphism $\rho:\cO_q\rightarrow{\rm End}(\cH)$, we often refer to $\cO_q$ as the \emph{operator algebra}, although elements in $\cO_q$ are functions on the phase space. Similarly, since $\rho(F)$ is an operator acting on $\cH$, we will often use the notation
\begin{equation}
\label{rho}
\rho(F)\equiv \hat F\in{\rm End}(\cH)\;,\quad\,F\in\cO_q\;.    
\end{equation}

The graded star commutators computed through \eqref{our star} are the usual ones imposed through canonical quantization, meaning that
\begin{align}
\label{bosonic comm}
    &[P_A,X^B]_\star=-i\hbar\,\delta_A^B\;.
\end{align}
In the following we will set $-i\hbar=1$, so that $[P_A,-]_\star=\del_{X^A}$ without phase factors. With this normalization, both gradings under $U(1)_\cG$ and $U(1)_\cJ$ are counted by the $\star$-vector fields $\cG$ and $\cJ$, respectively, generated by the functions
\begin{equation}\label{grading functions}
G:=cb+\cC\bar\cB+\bar\cC\cB \equiv c^ib_i\;,\quad J:=\alpha^\mu\bar\alpha_\mu +\cC\bar\cB+\cB\bar\cC\equiv\mathfrak{a}^m\bar{\mathfrak{a}}_m\;,  
\end{equation}
such that
\begin{equation}\label{grading vector fields}
\cG:=[G,-]_\star=c^i\frac{\del}{\del c^i}-b_i\frac{\del}{\del b_i}\;,\quad\cJ:=[J,-]_\star=\mathfrak{a}^m\frac{\del}{\del\mathfrak{a}^m}-\bar{\mathfrak{a}}_m\frac{\del}{\del\bar{\mathfrak{a}}_m} \;,   
\end{equation}
yield $\cG(c^i)=+c^i$, $\cG(b_i)=-b_i$ and $\cJ(\mathfrak{a}^m)=+\mathfrak{a}^m$, $\cJ(\bar{\mathfrak{a}}_m)=-\bar{\mathfrak{a}}_m$, as desired.

Moreover, since the BRST charge \eqref{Q} remains nilpotent in the quantized theory, i.e.~it obeys $Q\star Q\equiv\frac12\,[Q,Q]_\star=0$, we can endow both the operator algebra $\cO_q$ and the state space $\cH$ with differentials given by 
\begin{equation}\label{cal Q and hat Q}
\cQ:=[Q,-]_\star\in{\rm End}(\cO_q)\;,\quad \hat Q:=\rho(Q)\in{\rm End}(\cH)\;,    
\end{equation}
thus making $\cO_q$ into a differential graded algebra and $\cH$ into a chain complex.

To summarize, we have identified the state space $\cH$ of the bosonic spinning particle as the space $\cH=C^\infty(\mathbb{R}^{D})[\mathfrak{a}^m,c]$. We introduced the graded associative algebra $\cO_q=(\cO,\star)$ on the quantized phase space through a deformation of the pointwise product into a non-commutative star product. The algebra $\cO_q$ is additionally a Lie algebra under star commutator, which one can use to define an adjoint action on itself. Finally, using the appropriate representation \eqref{action on states}, we defined the action of $\cO_q$ on states in $\cH$. 

\subsection{Differential operators}

We now relate the above construction in terms of the star algebra of functions to differential operators, which are the more usual representation of abstract operator algebras in quantum mechanics.
The representations given in \eqref{rho normal} and \eqref{rho Weyl} can be viewed as maps 
\ba
\cO_q &\longrightarrow \text{End}(\cH) \; , \\
F &\longmapsto \hat F \; , 
\ea
where on the left hand side $\mathcal{O}_q$ is equipped with either $\star_{\rm n}$ or $\star_{\rm w}$. The formula for the star product shows that for a function $F \in \mathcal{O}_q$, the operator $\hat F \in \text{End}(\cH)$ is actually a differential operator on the algebra $\cH$. This means that $\hat F$ is of the form
\be
\hat F = \sum_{i = 0}^n  F^{A_1 \ldots A_i} \frac{\partial}{\partial X^{A_1}} \cdots \frac{\partial}{\partial X^{A_i}} \; .
\ee
Here, the first term in the sum (i.e.~$i = 0$) acts via left multiplication. The number $n$ is called the order of the operator $\hat F$, which is finite but can be arbitrarily large. The algebra of all differential operators on $\cH$ is called $\text{Diff}(\cH)$, and it is a subalgebra of $\text{End}(\cH)$.

Since any $F \in \mathcal{O}_q$ is polynomial in momenta, the map $F \mapsto \hat F$ is an isomorphism of associative algebras, given by the $\rho$ in  \eqref{rho}. This isomorphism is such that when expanding $\hat F$ in powers of $\hbar$
\be
\hat F = \sum_{k \ge 0} \hbar^k \hat F_k \; ,
\ee
the operator $\hat F_k$ is an order $k$ differential operator. 

We can think of a quantization procedure on $\cO$ as a choice of isomorphism of vector spaces $\cO \rightarrow \text{Diff}(\cH)$, with the condition that a function of the form $F^{A_1\ldots A_k}(X) P_{A_1} \cdots P_{A_k}$ gets mapped to
\be
F^{A_1\ldots A_k}(X) \frac{\partial}{\partial X^{A_1}} \cdots \frac{\partial}{\partial X^{A_k}} + \text{lower order terms}.
\ee
The isomorphism of vector spaces then induces a product $\star$ on $\cO$ induced by the composition product on $\text{Diff}(\cH)$. The products $\star_{\rm n}$ and $\star_{\rm w}$ could be derived in this way. For example, the isomorphism reproducing $\star_{\rm w}$ maps $x^\mu p_\nu$ to $x^\mu \partial_\nu + \tfrac{1}{2}\delta^\mu_\nu$, while $\star_n$ is reproduced by mapping $x^\mu p_\nu$ to $x^\mu \partial_\nu$.

We can also talk about the order of a differential operator with respect to $\star$. Although higher order operators with respect to non-commutative products can be defined \cite{ginzburg2010diffop}, for us it is important that the commutator $\cW := [W,-]_\star$, for $W \in \cO_q$, is a derivation of $\star$, but not necessarily of the pointwise product.

\section{State space decomposition and operator algebra quotients}

In this section, we decompose the state space $\cH$ of the bosonic spinning particle into sectors of different spin, focusing on the spin-1 sector $\cH_1$. Within the corresponding operator algebra, we identify an ideal $\mathfrak{I}$ under star product, which corresponds to the annihilator of the space $\cH_1$. In the coming sections, our goal will be to relate a special subalgebra of $\cO_q/\mathfrak{I}$ to the $C_\infty$ structure appearing in Yang-Mills theory upon color stripping.

\subsection{The spin-1 sector and its operator algebra}

We now introduce gradings for the state space $\cH$ associated with the $U(1)_\cG$ and $U(1)_{\cJ}$ symmetries of the theory. The grading by ghost number is implemented on $\cH$ by the operator 
\begin{equation}
\hat G:=\rho(G)=c\frac{\del}{\del c}+\cC\frac{\del}{\del\cC}-\cB\frac{\del}{\del\cB}\;.    
\end{equation}
We can decompose the space of states as the bounded direct sum $\cH=\oplus_{k=-1}^2\cH_k$, where elements $\psi_k$ of $\cH_k$ have homogeneous ghost number $\hat{G}=k$. We also define the operator $\hat  J$ associated with the $U(1)_\cJ$ charge as
\begin{equation}
    \hat J:=\rho(J)=\mathfrak{a}^m\frac{\del}{\del\mathfrak{a}^m}\;,
\end{equation}
which counts the number of $\mathfrak{a}^m$ modes in a state. This grading induces a decomposition for $\cH$ that is bounded from below, since the coordinates $X^A=(x^\mu,c,\mathfrak{a}^m)$ have positive or zero charge. We thus obtain
\begin{equation}
    \label{Hilbert space decomposition}
    \cH=\bigoplus_{j=0}^\infty\bigoplus_{k=-1}^2\cH_{j,k}\,.
\end{equation}

The differential on $\cH$ is $\hat Q$, which is nilpotent and has ghost number $+1$, as inherited from the charge $Q$. This means that $[\hat G,\hat Q]=\hat Q$. The differential is also neutral under $U(1)_\cJ$, since $[\hat{J},\hat{Q}]=0$. Thus, $\hat Q$ maps between sectors of the state space with ghost number differing by one, but with the same $U(1)_\cJ$ charge, as 
\begin{equation}
    \hat Q: \cH_{j,k}\to \cH_{j,k+1}\;.
\end{equation}
As such, $(\cH_j, \hat Q)$ is a chain complex for every $j$, with $\cH_j:=\oplus_{k=-1}^2\cH_{j,k}$. The physical content of its cohomology is a tower of free massless particles of spins ranging from $j$ to one or zero, decreasing in steps of 2, see e.g. \cite{Bonezzi:2024emt} for details. 

We now set $j=1$, limiting ourselves to the space $\cH_1$. This space is related to free Yang-Mills theory, since the cohomology of $(\cH_1,\hat Q)$ contains a single massless spin-1 particle. A general element of $\cH_1$ is a function $\psi(X)$ of the form
\begin{equation}
    \psi(X)=\psi_m(x,c)\,\mathfrak{a}^m\;.
\end{equation}

The operator algebra $\cO_q$ has a subalgebra of functions $F$ with $U(1)_\cJ$ charge equal to zero, given by the subspace $\ker\cJ\subset\cO_q$.
Under the map $\rho$, this subalgebra gives rise to diagonal operators in $\cH$, in that
\begin{equation}
\rho\,\rvert_{\ker\cJ}:\cH_j\rightarrow\cH_j\quad\forall\,j\;,    
\end{equation}
which, in particular, are endomorphisms of $\cH_1$. We recall from \eqref{grading vector fields} that
\begin{align}
\label{decomposition of J}
    \cJ=\mathfrak{a}^m\partial_{\mathfrak{a}^m}-\bar{\mathfrak{a}}_m\partial_{\bar{\mathfrak{a}}_m}=N_{\mathfrak{a}}-N_{\bar{\mathfrak{a}}}\;,
\end{align}
where we defined the separate number operators $N_{\mathfrak{a}}:=\mathfrak{a}^m\partial_{\mathfrak{a}^m}$ and $N_{\bar{\mathfrak{a}}}:=\bar{\mathfrak{a}}_m\partial_{\bar{\mathfrak{a}}_m}$.
Using this, we can write a general function $F$ in $\ker\cJ$ as
\begin{equation}\label{general j=0 operator}
\begin{split}
F&=\sum_{k=0}^\infty F_{(k)}\;,\quad N_{\mathfrak{a}}(F_{(k)})=N_{\bar{\mathfrak{a}}}(F_{(k)})=k\,F_{(k)}\;,   \\
F_{(k)}&=F_{m_1\dots m_k}^{ n_1\dots n_k}(x,p,c,b)\mathfrak{a}^{m_1}\dots \mathfrak{a}^{m_k} \mathfrak{\bar{a}}_{n_1}\dots\mathfrak{\bar{a}}_{n_k}\;.
\end{split}    
\end{equation}
A crucial remark is that any $\hat{F}_{(k)}\equiv\rho(F_{(k)})$ with $k>1$ annihilates any element of $\cH_1$, since any $\psi(X)\in\cH_1$ is linear in $\mathfrak{a}^m$, while
\begin{equation}
\hat F_{(k)}=\hat F_{m_1\dots m_k}^{ n_1\dots n_k}\mathfrak{a}^{m_1}\dots \mathfrak{a}^{m_k} \frac{\del}{\del\mathfrak{{a}}^{n_1}}\dots\frac{\del}{\del\mathfrak{{a}}^{n_k}}\;,    
\end{equation}
which can be derived from \eqref{our star} and \eqref{action on states}. We thus see that the annihilator of $\cH_1$ is isomorphic (through the map $\rho$) to the space of functions in $\ker\cJ$ with $k>1$, which we denote by $\mathfrak{I}$. Moreover, the subspace $\mathfrak{I}$ is a two-sided ideal of $\ker\cJ$, meaning that for every $F\in\ker\cJ$ one has
\begin{equation}
\mathfrak{I}\star F\subset\mathfrak{I}\;,\quad F\star\mathfrak{I}\subset\mathfrak{I} \;.  
\end{equation}
This can be proven as follows: given two functions $F_{(k_1)}$ and $F_{(k_2)}$ in $\ker\cJ$, with $N_{\bar{\mathfrak{a}}}=k_1$ and $N_{\bar{\mathfrak{a}}}=k_2$, respectively, their star product decomposes as
\begin{equation}\label{ideal proof}
F_{(k_1)}\star F_{(k_2)}=F_{(k_1+k_2)}+F_{(k_1+k_2-1)} +\dots+F_{({\rm max}(k_1,k_2))}\;,   
\end{equation}
so that, if either $k_1$ or $k_2$ is larger than 1, all terms in the right-hand side of \eqref{ideal proof} have $N_{\bar{\mathfrak{a}}}>1$ and thus belong to the ideal.

We are thus led to define the quotient space $\cR$ obtained by modding out the ideal from the space  $\ker{\cJ}\subset \cO_q$
\begin{equation}
    \label{R}
    \cR:= \faktor{\ker\cJ}{\mathfrak{I}}\;. 
\end{equation}
The space $\cR$ is thus effectively the space of operators acting nontrivially on $\cH_1$. Given that the elements of $\mathfrak{I}$ are the functions in $\ker\cJ$ with $N_{\bar{\mathfrak{a}}}>1$, one can define a projector that picks the obvious representatives in the equivalence classes in $\cR$:
\begin{align}
    \label{PR}
    &P_\cR\left(\sum_{k=0}^\infty F_{(k)}\right):=F_{(0)}+F_{(1)}\;.
\end{align}
Notice that the ideal is the kernel of this projector: $\mathfrak{I}=\ker P_\cR$. Although $\cR$ is defined as the quotient space \eqref{R}, we will practically work with the representatives \eqref{PR} and project again the results after taking the star product, which is equivalent to using a projected product $\star_\cR$ defined by
\begin{align}
    F\star_\cR G:=P_{\cR}(F\star G)\;,\quad \forall\,F,G\in{\rm Im}\,P_\cR\;,
\end{align}
so that $({\rm Im}\,P_\cR,\star_\cR)$ is also an associative algebra, but we will never write $P_\cR$ or $\star_\cR$ explicitly.

\subsection{Decomposition of the differential}

We now present in more detail the structure of the operator $\cQ=[Q,-]_\star$, which is the differential on the operator algebras $\cO_q$ and $\cR$. We first introduce the following decomposition of $\cQ$
\begin{align}
    \label{s decomposition}
    \cQ=\gamma+\delta\,,&&\gamma=d+\sigma+\Delta\,,
\end{align}
where we present its components explicitly as differential operators in $ {\rm Diff}(\cO_q)$, with respect to the pointwise product, as
\begin{align}
    &d:= C^\mu\partial_\mu\label{d}\,,\\
    &\sigma:= \cC \big(p\cdot\partial_\alpha+b\,\del_\cB\big)-\bar{\cC}\big(p\cdot\partial_{\bar{\alpha}}+b\,\del_{\bar\cB}\big)-\cM \partial_c\label{sigma}\,,\\
    &\Delta:=\frac{1}{2}\left(\cC \partial_{{\alpha}}^\mu +\bar{\cC}\partial_{\bar{\alpha}}^\mu\right)\del_\mu\label{Delta}\,,\\
    &\delta:=p^2\partial_b+\alpha\cdot p\,\partial_\cB+\bar\alpha\cdot p\,\partial_{\bar{\cB}}+\frac{1}{2}\left[ \alpha^\mu\partial_\mu\partial_\cB-\bar{\alpha}^\mu\partial_\mu\partial_{\bar{\cB}} -\cC\partial_{\cB}\partial_{c}-\bar{\cC}\partial_{\bar{\cB}}\partial_c \right]+\frac{1}{4}\B \partial_b\,,
\end{align}
where we defined
\begin{align}
\label{Cmu and M}
    &C^\mu:=\cC\bar{\alpha}^\mu+\bar{\cC}\alpha^\mu+2cp^\mu\,,&&\cM:=\cC\bar{\cC} \,.
\end{align}

The split of $\cQ$ into the operators $\gamma$ and $\delta$ comes from refining the ghost degree into pure ghost and pure antighost number, such that $\cG=N_{\rm gh}-N_{\rm agh}$. The ghost and antighost counting operators can be expressed as differential operators in the following way
\begin{align}
    N_{\rm gh}:=c^i\frac{\del}{\del c^i}\,,&& N_{\rm agh}:=b_i\frac{\del}{\del b_i}\,,
\end{align}
in agreement with \eqref{grading vector fields}, where we imply summation over the ghost and antighost multiplets respectively. This refinement of the ghost degree assigns
\begin{align}
    \label{gamma delta pure ghost antighost}
    &N_{\rm gh}(\gamma)=+1\,,&&N_{\rm agh}(\gamma)=0\,,\notag \\
    &N_{\rm gh}(\delta)=0\,,&&N_{\rm agh}(\delta)=-1\,.
\end{align}
Therefore, the split is induced by either pure ghost or pure antighost number. Expanding the $\cQ^2=0$ condition into
\begin{equation}
0=\gamma^2+\delta^2+[\gamma,\delta]\;,    
\end{equation}
we see that the three terms above have, for instance, pure ghost number $+2$, $0$ and $+1$, respectively, and must thus vanish separately. This implies that both $\gamma$ and $\delta$ are nilpotent differentials that commute with each other: $\gamma^2=\delta^2=[\gamma,\delta]=0$. 

The introduction of pure ghost and antighost number grading leads us to a decomposition of $\cO_q$, depending on whether operators contain antighost variables:
\begin{equation}\label{hilbert space decomposition b}
\begin{split}
\cO_q&\equiv\cO_{q,b}\oplus\cO_{q,c}\;,\\
\cO_{q,b}&:=\{F \in \cO_q|N_{\rm agh}(F)>0\}\;,\quad \cO_{q,c}:=\{F\in \cO_q|N_{\rm agh}(F)=0\}\;.
\end{split}    
\end{equation}
While the subspace $\cO_{q,b}$ is \emph{not} closed under star product, the space $\cO_{q,c}$ is, and forms a subalgebra. The differential on $\cO_{q,c}$ is $\gamma$, in that $\cQ(\cO_{q,c})\equiv \gamma(\cO_{q,c})$.
 The pure ghost number grading is inherited by the subspaces $\ker{\cJ}$ as well as $\cR$, leading to the spaces $\ker{\cJ}_{c}$ and $\cR_{c}$. As for the star product, when performing computations in $\cR$ or $\cR_c$ we always pick the representatives \eqref{PR} but omit the projector $P_\cR$, implying that all results are presented up to elements in the ideal $\mathfrak{I}$.

The differential $\gamma$ can be further split into $d+\sigma$, which are first-order differential operators, and $\Delta$, which is second order. We have separated $d$ and $\sigma$ based on their derivative structure: $d$ contains spacetime derivatives $\frac{\del}{\del x^\mu}$, while $\sigma$ does not. One can further verify that these operators are themselves nilpotent differentials: $d^2=\sigma^2=\Delta^2=0$ and that they commute with one another.


\section{Homotopy transfer to the $C_\infty$ algebra of Yang-Mills}

In this section, we briefly review the $C_\infty$ algebra of Yang-Mills theory and its relation to the spin-one sector of the state space $\cH_1$. Elements of the algebra are generated from the vacuum of $\cH_1$ through the action of vertex operators as presented in \cite{Bonezzi:2024fhd}. The functions corresponding to these operators define the algebra $(\cF,\cQ,\star)$, a strict differential, associative and non-commutative algebra which contains all Yang-Mills vertex operators and is homotopy equivalent to the $C_\infty$ algebra of the theory.

\subsection{A subalgebra for Yang-Mills vertex operators}

Perturbative Yang-Mills theory, upon stripping off color degrees of freedom, exhibits a $C_\infty$ algebra structure \cite{Zeitlin:2008cc,Borsten:2021hua,Bonezzi:2022yuh}. A $C_\infty$ algebra consists of a graded vector space $\cK$ endowed with a collection of multilinear maps $m_n:\cK^{\otimes n}\to \cK$. It is a generalization of a differential graded commutative algebra, where the map $m_1$ serves as a differential and the bilinear product $m_2$ is graded commutative, but associative only up to homotopy. 

In the case of Yang-Mills theory, the underlying vector space is isomorphic to a suspension (degree shift) of the state space $\cH_1$:
\begin{align}
    \label{K}
    \cK=\bigoplus_{i=0}^3\cK_i\,,&&\cK_i:=\cH_{1,i-1}\,.
\end{align}
As explained in more detail in \cite{Bonezzi:2024fhd}, elements of $\cK$ are given as states in $\cH_1$ by 
\begin{align}
    &\lambda:=\lambda(x)\,\cB\ \in \ \cK_0\,,\label{lambda in K}\\
    &\cA:=A_\mu(x)\,\alpha^\mu+\varphi(x)\,c\cB\ \in \ \cK_1\,,\label{cA in K}\\
    &\cE:=E_\mu(x)\,c\,\alpha^\mu+E(x)\,\cC \ \in \ \cK_2\,,\label{cE in K}\\
    &\cN:=N(x)\,c\,\cC\ \in \ \cK_3\,,\label{cN in K}
\end{align}
which corresponds to the Yang-Mills theory for a gauge boson $A_\mu$ equipped with an auxiliary scalar field $\varphi$ as presented in \cite{Bonezzi:2022yuh,Bonezzi:2024emt}. In the following, we will denote generic elements of $\cK$ as $\psi$, intended as functions $\psi(X)\in\cH_1$.

The differential map $m_1$ for this algebra is given through the action of the BRST operator on states in $\cH_1$ as
\begin{align}
\label{m1}
    m_1\equiv\hat Q=c\,\B+\left(\cC\,\frac{\del}{\del\alpha_\mu}+\alpha^\mu\,\frac{\del}{\del\cB}\right)\del_\mu-\cC\,\frac{\del}{\del\cB}\frac{\del}{\del c}\;,
\end{align}
which is derived from the map $\rho$ \eqref{action on states}.
Therefore, color-stripped Yang-Mills theory is encoded in the chain complex
\begin{center}
    \begin{tikzpicture}
    \node at (-2,0) {$0$};
    \draw[->] (-1.5,0)--(-0.5,0);
        \node at (0,0) {$\cK_{0}$};
        \draw[->] (0.5,0)--(1.5,0);
        \node at (1,0.3) {$m_1$};
        \node at (2,0) {$\cK_{1}$};
        \draw[->] (2.5,0)--(3.5,0);
        \node at (3,0.3) {$m_1$};
        \node at (4,0) {$\cK_{2}$};
        \draw[->] (4.5,0)--(5.5,0);
        \node at (5,0.3) {$m_1$};
        \node at (6,0) {$\cK_{3}$};
        \draw[->] (6.5,0)--(7.5,0);
        \node at (7,0.3) {};
        \node at (8,0) {$0$};
        \node at (0,-1) {$\lambda$};
        \node at (2,-1) {$\cA$};
        \node at (4,-1) {$\cE$};
        \node at (6,-1) {$\cN$};
    \end{tikzpicture}
\end{center}
The physical meaning of this chain complex is the following: the fields of our theory, $A_\mu$ and $\varphi$, are contained in the element $\cA\in\cK_1$, at ghost number 0. At ghost number $-1$, so in the space $\cK_0$, we find gauge parameters, which induce gauge transformations given as $m_1(\lambda)$. The action of the differential on fields results in (linearized) equations of motion, which live in the space $\cK_2$ and have the form \eqref{cE in K}. Finally, the gauge symmetry leads to relations between the equations of motion themselves, the Noether identities, which are given by $m_1(\cE)$ and will be elements of $\cK_3$ of the form \eqref{cN in K}. We further assign target space mass dimension to $\{\lambda, A_\mu, \varphi, E, E_\mu, N\}$ as follows: $[\lambda]=0$, $[A_\mu]=1$, $[\varphi]=2$, $[E_\mu]=3$, $[E]=2$, $[N]=4$. The absence of mass parameters in the theory helps in controlling locality of all the maps to be defined in the following.

A different point of view on this algebra was introduced in \cite{Bonezzi:2024fhd}. The space $\cH_1$, hence $\cK$, contains a ``vacuum state'' represented by the function $\psi_{\rm vac}(X)=\cB$. This state is interpreted as a constant gauge parameter in $\cK_0$ and obeys $m_1(\psi_{\rm vac})=0$. As a closed element of degree 0, it belongs to the cohomology of $m_1$. Arbitrary elements of $\cK$ can then be generated from the vacuum by applying suitable vertex operators, in the sense
\begin{equation}
    \label{bacuum}
    \psi=\rho(V(\psi);\cB)= \hat{V}(\psi)\psi_{\rm vac}\in\cH_1\,,
\end{equation}
where $V(\psi)$ is a function in $\cR_c\subset \cO_q$. Since the vacuum $\psi_{\rm vac}$ belongs to $\cH_1$, the vertex operators must be elements of $\ker\cJ$ and have $\cG(V(\psi))=|\psi|$, where $|\cdot|$ is the degree in $\cK$.
We further demand $V$ to be a chain map between $({\rm End}(\cK),m_1)$ and $(\cR_c,\cQ)$, meaning that, for $\psi\in \cK$
\begin{equation}\label{V chain}
    \cQ \big(V(\psi)\big)=V\big(m_1(\psi)\big)\,.
\end{equation}
These requirements lead to the following functions $V(\psi)$ for elements of $\cK$ of different degrees:
\begin{equation}\label{V of psi}
\begin{split}
V(\lambda)&=\lambda(x)\;,\\
    V(\cA)&=C^\mu A_\mu(x)+C^{\mu\nu}f_{\mu\nu}(x)+c\,\big(\varphi(x)-\del^\mu A_\mu(x)\big)\;,\\
    V(\cE)&=\cM E(x)+\cM^\mu \partial_\mu E(x)+\tilde{\cM}^\mu(E_\mu-\partial_\mu E)\;,\\
    V(\cN)&=eN(x)\;,    
\end{split}    
\end{equation}
where $f_{\mu\nu}=\del_\mu A_\nu-\del_\nu A_\mu$ is the Abelian field strength, and we introduced the following notation for specific elements of $\cR_c$, 
\begin{align}
    C^{\mu\nu}&:=2c\,\alpha^{[\mu}\bar{\alpha}^{\nu]}\label{calM}\,,\\
    \cM^\mu&:=c\,(\bar{\cC}\alpha^\mu+\cC\bar{\alpha}^\mu)\,,\label{Mmu}\\
    \tilde{\cM}^\mu&:= c\,(\bar{\cC}\alpha^\mu-\cC\bar{\alpha}^\mu)\,,\label{Mtilde}\\
    e&:=c\,\cC\bar{\cC}\,.\label{e}
\end{align}
As remarked, all of these vertex operators belong to $\cR_c\subset \cR$, meaning that they contain no dependence on the $b_i$, equivalently $N_{\rm agh}V(\psi)=0$ for all $\psi\in \cK$. Moreover, all of their $x$-dependence is given through  polynomials in $\{c^i,\alpha^\mu,\bar{\alpha}^\mu,p_\mu\}$ with coefficients in $C^\infty(\mathbb{R}^D)$. Given the target space mass dimensions, all $V(\psi)$ have total mass dimension zero.

These vertex operators do not form a closed subalgebra of $(\cR_c,\cQ,\star)$. We now seek their minimal extension to a set of operators that create elements of $\cK$ from the vacuum $\psi_{\rm vac}$, while simultaneously forming a closed algebra under $\star$ and $\cQ$. We identify the following such elements within $\cR_c$:
\begin{align}
    &F_0=\lambda(x)\,,\label{F_0}\\
    &F_1=C^\mu A_\mu(x)+C^{\mu\nu }B_{\mu\nu}(x)+c\,\pi(x)\,,\label{F_1}\\
    &F_2=\cM\, w(x)+\cM^\mu w_\mu(x)+\tilde{\cM}^\mu \tilde{w}_\mu(x) +\cM^{\mu\nu}w_{\mu\nu}(x)\,,\label{F2}\\
    &F_3=eN(x)+e^\mu N_\mu(x)\,,\label{F3}
\end{align}
where we further defined
\begin{align}
    \cM^{\mu\nu}&:=\frac{1}{2}C^\mu C^\nu=2c\,p^{[\mu}(\bar\cC\alpha^{\nu]}+\cC\bar\alpha^{\nu]})\,,\label{cM munu}\\ e^\mu&:=e p^\mu\,.\label{ep}
\end{align}
These elements span a space $\cF:=C^\infty(\mathbb{R}^D)[B_\cF]$, where $B_\cF$ is the following basis:
\begin{align}
    \label{basis of F}
    B_\cF:=\left\{1;c,C^\mu,C^{\mu\nu};\cM,\cM^\mu,\tilde{\cM}^\mu,\cM^{\mu\nu};e,e^\mu\right\}\,.
\end{align}
The space $\cF$ inherits the grading of $\cR_c\subset\cO_q$ and we have $\cF=\oplus_{i=0}^3\cF_i$, with the $F_i$ of \eqref{F_0}-\eqref{F3} belonging to the space $\cF_i$.

We can now check that $\cF$ is closed under $\star$ by computing the products at each degree to be
\begin{align}
&F_{0,1}\star F_{0,2}=\lambda_1\lambda_2\,,\label{F0*F0}\\
    &F_0\star F_1=C^\mu(\lambda A_\mu)+C^{\mu\nu}(\lambda B_{\mu\nu})+c(\pi-A\cdot\partial)\lambda\,,\label{F0*F1}\\
    &F_0\star F_2=\cM(\lambda w)+\cM^\mu( w_\mu+w_{\mu\nu}\partial^\nu) \lambda+\tilde{\cM}^\mu(\lambda\tilde{w}_\mu)+\cM^{\mu\nu}(\lambda w_{\mu\nu})\,,\label{F0*F2}\\
    &F_0\star F_3=e( N\lambda-\frac{1}{2}N^\mu\partial_\mu\lambda)+e^\mu(\lambda N_\mu) \label{F0*F3}\,,\\
    &F_{1,1}\star F_{1,2}=\cM A_1\cdot A_2+2\cM^\mu\left[A_{(1}\cdot\partial A_{2)\mu}-A_{(1}^\nu B_{2)\nu\mu}+\pi_{[1} A_{2]
    \mu}\right]\notag \\&+2\cM^{\mu\nu}A_{1\mu}A_{2\nu}+2\tilde{\cM}^\mu A^\nu_{[1}B_{2]\nu\mu}\,,\label{F1*F1}\\
    &F_1\star F_2=e\Big[(\pi+A\cdot\partial)w-A_\mu(w^\mu+\tilde{w}^\mu)+\partial^\mu A^\nu w_{\mu\nu}\Big]+2e^\mu(A_\mu w-w_{\mu\nu}A^\nu)\,.\label{F1*F2}
\end{align}
The star product in $\cF$ is still not commutative, with graded commutators given by
\begin{align}
    &[F_0,F_1]_\star=-2cA\cdot \partial \lambda\,,\label{[F0,F1]}\\
    &[F_0,F_2]_\star=2\cM^{\mu}w_{\mu\nu}\partial^\nu \lambda \label{[F0,F2]}\,,\\
    &[F_0,F_3]_\star=-eN^\mu\partial_\mu\lambda\,,\label{[F0,F3]}\\
    &[F_{1,1},F_{1,2}]_\star=2\cM A_1\cdot A_2+4\cM^\mu \left(A_{(1}\partial \cdot A_{2)\mu}-A_{(1}^\nu B_{2)\nu\mu}\right) \,,\\
    &[F_1,F_2]_\star=2e\,A^\mu(\del_\mu w-w_\mu)-2e^\mu w_{\mu\nu}A^\nu\,.\label{[F1,F2]}
\end{align}
We further verify that $\cF$ is closed under the action of the differential $\cQ$:
\begin{align}
    &\cQ(F_0)=C^\mu\partial_\mu\lambda\,,\label{sF0}\\
    &\cQ(F_1)=-\cM\pi-\cM^\mu\partial_\mu \pi-\tilde{\cM}^\mu\partial^\nu B_{\mu\nu}+\cM^{\mu\nu}(f_{\mu\nu}-B_{\mu\nu})\,,\label{sF1}\\
    &\cQ(F_2)=2e^\mu(\partial_\mu w-w_\mu)-e\,\partial^\mu \tilde{w}_\mu\label{sF2}\;.
\end{align}
In conclusion, $(\cF,\cQ,\star)$ defines a differential graded associative algebra. Since all basis elements of $\cF$ are in $\cR_c$, the effective differential on the space is $\gamma$.
We remark that, if one would interpret the $x$-dependent functions at degree +1 as the field content of a theory, it would be a theory with a vector field $A_\mu$, as well as a scalar field $\pi$ and two-form $B_{\mu\nu}$ as auxiliary fields. This suggests  that the minimal extension of the algebra of worldline vertex operators \eqref{V of psi} that is closed under $\cQ$ and $\star$ is related to a generalization of first order Yang-Mills theory. 

We will now show how the basis elements of $\cF$ are connected to the cohomology of the differential $\sigma$, and that the latter encodes the ``spectrum'' of the  physical sector of the theory. Since $\sigma$ is insensitive to the $x-$dependence, it is sufficient to compute it on the span of the basis $\mathbb{C}[B_\cF]$ with $x$-independent coefficients, yielding
\begin{equation}
\begin{split}
\sigma(1)&=0\;,\\
\sigma(C^\mu)&=0\;,\quad\sigma(C^{\mu\nu})=-\cM^{\mu\nu}\;,\quad\sigma(c)=-\cM\;,\\
\sigma(\tilde{\cM}^\mu)&=0\;,\quad \sigma(\cM^\mu)=-2\,e^\mu\;,\\
\sigma(e)&=0\;,
\end{split}    
\end{equation}
where we omitted $\sigma$ on exact elements, such as $\cM$ and $\cM^{\mu\nu}$, since it is obviously zero.
It turns out that the elements of $B_\cF$ split into a subspace in the $\sigma$-cohomology, generated by $\{1,C^\mu,\tilde{\cM}^\mu,e\}$, plus the trivial pairs
\begin{align}
    \label{trivial pairs}
    &\sigma(c)+\cM=0\,,&\sigma(\cM^\mu)+2\,e^\mu=0\,,&&\sigma(C^{\mu\nu})+\cM^{\mu\nu}=0\,.
\end{align}

We can now make the observation that physical objects, meaning $A_\mu$ with its gauge parameter, together with the corresponding equation of motion and Noether identity, appear with a basis element that is in cohomology. On the contrary, the auxiliary fields $\pi$ and $B_{\mu\nu}$ and their associated elements in other degrees appear with basis elements that form trivial pairs. The underlying reason is that the auxiliary scalar field and two-form can be integrated out from the theory in a local way, without altering its physical content. In fact, in the next subsection we will relate the $C_\infty$ algebra of color stripped Yang-Mills theory to the star algebra of $\cF$ via a local homotopy transfer.



\subsection{Homotopy transfer from $\cF$ to $\cK$}
We will now perform homotopy transfer between the chain complexes $(\cF,\cQ)$ and $(\cK,m_1)$ by constructing a quasi-isomorphism between the two algebras. Such a map is the natural morphism between homotopy algebras, the existence of which guarantees that the algebras have isomorphic cohomologies. In physics terms, these algebras describe theories with an equivalent physical content at the perturbative level. For more details on homotopy transfer, we refer the reader to \cite{Arvanitakis:2020rrk,Arvanitakis:2021ecw,Chiaffrino:2020akd}.

To perform homotopy transfer between the chain complexes, we need to construct the diagram below.

\begin{figure}[h]
\begin{centering}
\label{fig:homotopy transfer}
    \begin{tikzpicture}
        \node at (0,0) {$\cF:$};
        \node at (1,0) {$0$};
        \draw[->] (1.5,0)--(2.5,0);
        \node at (3,0) {$\cF_0$}; 
        \draw[->] (3.5,0)--(4.5,0);
        \node at (4,0.3) {$\cQ$};
        \node at (5,0) {$\cF_1$}; 
        \draw[->] (5.5,0)--(6.5,0);
        \node at (6,0.3) {$\cQ$};
        \node at (7,0) {$\cF_2$}; 
        \draw[->] (7.5,0)--(8.5,0);
        \node at (8,0.3) {$\cQ$};
        \node at (9,0) {$\cF_3$}; 
        \draw[->] (9.5,0)--(10.5,0);
        \node at (10,0.3) {$\cQ$};
        \node at (11,0) {$0$}; 
        \node at (0,-2) {$\cK:$};
        \node at (1,-2) {$0$};
        \draw[->] (1.5,-2)--(2.5,-2);
        \node at (3,-2) {$\cK_0$}; 
        \draw[->] (3.5,-2)--(4.5,-2);
        \node at (4,-1.7) {$m_1$};
        \node at (5,-2) {$\cK_1$}; 
        \draw[->] (5.5,-2)--(6.5,-2);
        \node at (6,-1.7) {$m_1$};
        \node at (7,-2) {$\cK_2$}; 
        \draw[->] (7.5,-2)--(8.5,-2);
        \node at (8,-1.7) {$m_1$};
        \node at (9,-2) {$\cK_3$}; 
        \draw[->] (9.5,-2)--(10.5,-2);
        \node at (10,-1.7) {$m_1$};
        \node at (11,-2) {$0$}; 
        \draw[->] (2.8,-0.3)--(2.8,-1.7);
        \node at (2.5,-1) {$\Pi_0$};
        \draw[->] (3.2,-1.7)--(3.2,-0.3);
        \node at (3.5,-1) {$I_0$};
        \draw[->] (4.8,-0.3)--(4.8,-1.7);
        \node at (4.5,-1) {$\Pi_1$};
        \draw[->] (5.2,-1.7)--(5.2,-0.3);
        \node at (5.5,-1) {$I_1$};
        \draw[->] (6.8,-0.3)--(6.8,-1.7);
        \node at (6.5,-1) {$\Pi_2$};
        \draw[->] (7.2,-1.7)--(7.2,-0.3);
        \node at (7.5,-1) {$I_2$};
        \draw[->] (8.8,-0.3)--(8.8,-1.7);
        \node at (8.5,-1) {$\Pi_3$};
        \draw[->] (9.2,-1.7)--(9.2,-0.3);
        \node at (9.5,-1) {$I_3$};
        \draw[->] (5,0.3) to [bend right=45] (3,0.3);
        \node at (4,1) {$H_1$};
        \draw[->] (7,0.3) to [bend right=45] (5,0.3);
        \node at (6,1) {$H_2$};
        \draw[->] (9,0.3) to [bend right=45] (7,0.3);
        \node at (8,1) {$H_3$};
    \end{tikzpicture}
    \caption{Homotopy transfer from $\cF$ to $\cK$.}
\end{centering}
    \end{figure}
In essence, we need to first introduce a projection $\Pi:\cF\to\cK$, as well as an inclusion $I: \cK\to\cF$ which are chain maps between their differentials:
\begin{align}
    \label{chain maps}
    \Pi\circ \cQ=m_1\circ\Pi\,,&& \cQ\circ I=I\circ m_1\,.
\end{align}
The projection and inclusion maps are then required to be inverting each other up to corrections by a map $H:\cF\to \cF$, meaning that
\begin{align}
    \label{quasi iso relation}
    \Pi\circ I=1_{\cK}\,, &&I\circ \Pi=1_\cF+\cQ H+H\cQ\,.
\end{align}
If this holds, the homotopy transfer theorem ensures that if $(\cF,\star)$ has a homotopy associative (or $A_\infty$) structure, then so should $(\cK,m_1)$, with higher maps $\mu_n$ that can be constructed using the homotopy data $(\Pi, I, H)$. Note that in this case, $(\cF,\star)$ is a strict associative algebra, meaning it is a special case of an $A_\infty$ algebra where all products of arity greater than two vanish. Even so, the resulting algebra after homotopy transfer is expected to be only associative up to homotopy.

For the projector $\Pi:\cF\to\cK\simeq\cH_1$, we take the action of the operator $\rho(F)$ on the vacuum:
\begin{equation}
\Pi:\,\cF\,\rightarrow\,\cK \;,\quad \Pi(F):=\rho(F;\psi_{\rm vac})\;,  
\end{equation}
where $\rho$ is the representation of $(\cO_q,\star)$ on states in $\cH$ as defined in \eqref{action on states} and we recall that $\psi_{\rm vac}(X)=\cB$.
To verify that the chain map condition holds we first compute
\begin{equation}\label{chain map P1}
\begin{split}
m_1\big(\Pi(F)\big)&=\rho\big(Q;\Pi(F)\big)=\rho\big(Q;\rho(F;\psi_{\rm vac})\big)\\
&=\rho(Q\star F;\psi_{\rm vac})\;,
\end{split}    
\end{equation}
where we used the algebra morphism property \eqref{representation rho}. We now write $Q\star F=[Q,F]_\star+(-1)^FF\star Q$ and make use of \eqref{representation rho} again to obtain
\begin{equation}\label{chain map P2}
\begin{split}
\rho(Q\star F;\psi_{\rm vac})&=\rho([Q,F]_\star;\psi_{\rm vac})+(-1)^F\rho\big(F;\rho(Q;\psi_{\rm vac})\big)\\
&=\rho(\cQ(F);\psi_{\rm vac})=\Pi\big(\cQ(F)\big)\;,
\end{split}    
\end{equation}
which concludes the proof, upon recognizing that $\rho(Q;\psi_{\rm vac})\equiv m_1(\psi_{\rm vac})=0$.
The explicit action of the projection on the general elements $F_i$ of $\cF$ is
\begin{align}
    &\Pi(F_0)=\lambda\,\cB\,,\label{P(F0)}\\
    &\Pi(F_1)=A_\mu \alpha^\mu +c\,\cB\,(\pi+\partial\cdot A)\,,\label{P(F1)}\\
    &\Pi(F_2)=c\,\alpha^\mu(w_\mu+\tilde{w}_\mu-\partial^\nu w_{\mu\nu})+\cC w\,,\label{P(F2)}\\
    &\Pi(F_3)=c\,\cC(N+\frac{1}{2}\partial^\mu N_\mu)\,.\label{P(F3)}
\end{align}
On the other hand, the natural way to include elements of $\cK$ as given in \eqref{lambda in K}-\eqref{cN in K} in $\cF$ is through the corresponding vertex operators, so we have $I\equiv V$, with $V(\psi)$ given in \eqref{V of psi}, and the chain map condition for $I$ is the property \eqref{V chain}.

Coming to the homotopy map, the defining relation \eqref{quasi iso relation} implies that it needs to have ghost number $-1$ and $U(1)_\cJ$ charge zero. We also require mass dimension $[H]=0$, since this is the case for all functions $V(\psi)$ as well as for the differential $\cQ$. Since this map is in ${\rm End}(\cF)$, it is not expected to depend on any of the $b_i$. We identify the map
\begin{equation}
    \label{homotopy}H:=\bar{\alpha}\cdot\partial_p\partial_{\bar{\cC}}\;,
\end{equation}
which fulfills all previously listed requirements and satisfies \eqref{quasi iso relation}, as can be shown by direct computation
\begin{equation}\label{check quasi iso}
\begin{split}
&(I\Pi-1) F_0=0=H \cQ F_0\,,\\
    &(I\Pi-1) F_1=C^{\mu\nu}(f_{\mu\nu}-B_{\mu\nu})=(H\cQ + \cQ H)F_1\,,\\
    &(I\Pi-1) F_2=\cM^\mu (\partial_\mu w-w_\mu)-\tilde{\cM}^\mu(\partial_\mu w-w_\mu+\partial^\nu w_{\mu\nu})-\cM^{\mu\nu}w_{\mu\nu}=(H\cQ + \cQ H)F_2\,,\\
    &(I\Pi-1) F_3 =\tfrac12\,e\,\del^\mu N_\mu-e^\mu N_\mu=\cQ HF_3\,.    
\end{split}    
\end{equation}
The homotopy $H$ acts trivially on the $x$-dependent part of elements of $\cF$, but relates different basis elements as 
\begin{align}
    \label{homotopy on basis elements}
H(\cM^{\mu\nu})=C^{\mu\nu}\,,&&H(e^\mu)=\tfrac12\,(\cM^\mu-\tilde{\cM}^\mu)\,.
\end{align}
Essentially, comparing with the trivial pairs \eqref{trivial pairs}, we see that $H$ is the map that inverts the action of $\sigma$ on the specific basis elements that appear along with the auxiliary two-form and its related objects in other degrees of the chain complex. As a result, $H$ implements the process of integrating out of $B_{\mu\nu}$.

We have thus verified that the homotopy data $(\Pi,I,H)$ define a quasi-isomorphism between the chain complexes $(\cF, \cQ)$ and $(\cK,m_1)$. Note that the choice of homotopy is in general not unique, however the map we present seems to be the minimal choice achieving the desired homotopy transfer.
We now use the homotopy data to construct the 2-product of $(\cK,m_1)$ out of the original product $\star$. Note that, since the star product is associative, but not commutative, we do not expect the transferred product to be the one of the $C_\infty$ algebra of Yang-Mills (as presented for example in \cite{Bonezzi:2024fhd}), but an $A_\infty$ product $\mu_2$, given by
\begin{equation}
    \label{transferred product}
    \mu_2:=\Pi\circ\star\circ (I\otimes I)\;.
    \end{equation}
The explicit form of this product on different sectors of $\cK$ is the following
\begin{align}
    &\mu_2(\lambda_1,\lambda_2)=\lambda_1 \lambda_2\,,\label{mu2 lambda lambda}\\
    &\mu_2(\lambda,\cA)=(\lambda A_\mu)\alpha^\mu+(\lambda\varphi)c\cB=\mu_2(\cA,\lambda)-2\,A\cdot \partial \lambda\,c\,\cB\;,\label{mu2 lambda A}\\
    &\mu_2(\lambda, \cE)=(\lambda E_\mu)c\alpha^\mu+\lambda E \cC=\mu_2(\cE,\lambda)\,,\label{mu2 lambda E}\\
    &\mu_2(\lambda, \cN)=(\lambda N)c\,\cC=\mu_2(\cN,\lambda)\,.\\
    &\mu_2(\cA_1,\cA_2)=(A_1\cdot A_2)\,\cC+2\left(\varphi_{[1}A_{2]\mu}+2A_{[1}\cdot \partial A_{2]\mu}+\del_\mu A_1\cdot A_2\right)c\alpha^\mu\,,\label{mu2 A1 A2}\\
    & \mu_2(\cA,\cE)=(\varphi E+2 A\cdot \partial E-A^\mu E_\mu)c\,\cC=\mu_2(\cE,\cA)\,,\label{mu2 A E}
\end{align}
and it is identical to the $\mu_2$ computed in \cite{Bonezzi:2024fhd} from the composition of two vertex operators acting on the vacuum state. Applying the inclusion map $I\equiv V$ as given in \eqref{V of psi} to the products $\mu_2$ of two elements $\psi_1$, $\psi_2\in \cK$, we find that
\begin{equation}
    \label{inclusion of mu2}
    V(\mu_2(\psi_1,\psi_2))=V(\psi_1)\star V(\psi_2)+\partial V_2(\psi_1,\psi_2)\,,
\end{equation}
where we defined
\begin{equation}
\begin{split}
V_2(\psi_1,\psi_2)&:=H\big(V(\psi_1)\star V(\psi_2)\big)\;,\\
\partial V_2(\psi_1,\psi_2)&:=\cQ V_2(\psi_1,\psi_2)+V_2(m_1(\psi_1),\psi_2)+(-1)^{\psi_1}V_2(\psi_1,m_1(\psi_2))\;.
\end{split}    
\end{equation}
This bilinear map $V_2$ coincides with the bilinear vertex operator which was defined in \cite{Bonezzi:2024fhd}. We now showed that this object is derived through the homotopy transfer of $(\cF,\cQ,\star)$ to $(\cK, m_1, \mu_2)$.

In \cite{Bonezzi:2024fhd}, it was also shown that $\mu_2$ can be mapped to a  \emph{commutative} product $m_2$ on $\cK$ via
\begin{equation}
    \label{mu2 to m2}
    m_2(\psi_1,\psi_2):=\mu_2(\psi_1,\psi_2)+m_1(f_2(\psi_1,\psi_2))+f_2(m_1(\psi_1),\psi_2)+(-1)^{\psi_1}f_2(\psi_1,m_1(\psi_2))\,.
\end{equation}
Since the shift is exact, the two products are homotopy equivalent, with homotopy $f_2$ given in \cite{Bonezzi:2024fhd}. In fact, ${\rm id}+f_2$ is an isomorphism of $A_\infty$ algebras, so that $(\cK,m_2,\dots)$ is a $C_\infty$ algebra, the one which describes color stripped Yang-Mills theory.
We can show that the relation between $\mu_2$ and the commutative $m_2$ in $\cK$ also descends from an analogous relation in $\cF$: to this end, consider the symmetrization of the star product
\begin{equation}
\label{M2}
M_2(F_1,F_2):=\frac12\,\Big(F_1\star F_2+(-1)^{F_1F_2}F_2\star F_1\Big)   \;.
\end{equation}
One can construct a degree $-1$ graded antisymmetric map $N_2:\cF^{\otimes2}\to\cF$ such that
\begin{align}
    M_2(F_1,F_2)=F_1\star F_2+\cQ N_2 (F_1,F_2)+N_2(\cQ F_1,F_2)+(-1)^{F_1}N_2(F_1,\cQ F_2)\,,
\end{align}
meaning that the star product on $\cF$ is commutative up to the homotopy $N_2$. On the other hand, the product $M_2$ is only associative up to homotopy and ${\rm id}+N_2$ is an isomorphism of $A_\infty$ algebras.
The general form of this map is
\begin{equation}
\label{N2}
N_2(F_1,F_2)=\tfrac14\,\del_c\del_p F_1\cdot\del_p F_2 \;,  \end{equation}
with the following explicit form at each degree
\begin{equation}
\begin{split}
N_2(F_{1,1},F_{1,2})&=c\,A_1\cdot A_2\;,\\
N_2(F_1,F_2)&=\cM^\mu A^\rho w_{\rho\mu}\;,\\
N_2(F_1,F_3)&=\tfrac12\,e\,A^\mu N_\mu\;.
\end{split}    
\end{equation}
The homotopy for commutativity $N_2$ can then be projected to $\cK$, yielding $f_2$ as
\begin{equation}
    \label{transferred f2}
    f_2=\Pi\circ N_2\circ (I\otimes I)\;.
    \end{equation}

In conclusion, we have established the following relations between our different algebras:

\begin{center}
    \begin{tikzpicture}
        \node at (0,0) {$(\cF,\cQ,\star)$};
        \node at (4,0) {$(\cF,\cQ,M_2)$};
        \node at (0,-3) {$(\cK, m_1,\mu_2)$};
        \node at (4,-3) {$(\cK, m_1, m_2)$};
        \draw[->](0.9,0)--(3,0);
        \draw[->] (0,-0.5)--(0,-2.7);
        \draw[->] (1,-3)--(2.9,-3);
        \draw[->] (4,-0.5)--(4,-2.7);
        \node at (2,0.5) {$N_2$};
        \node at (-1.3,-1.5) {$(\Pi,I,H)$};
        \node at (2,-3.5) {$f_2$};
    \end{tikzpicture}
\end{center}
In this diagram, horizontal arrows are homotopy equivalences between products, while vertical maps represent quasi-isomorphisms under homotopy transfer. The rightmost arrow indeed represents a homotopy transfer akin to the one we presented in this section, mapping between $C_\infty$ algebras in that case.

\section{Outlook}

In this paper, we have established that the $C_\infty$ algebra of color-stripped Yang-Mills theory is homotopy equivalent to a strict algebra of quantum worldline operators. In doing so, we have clarified how the nonlinear vertex operators introduced in \cite{Bonezzi:2024fhd} arise from the standard formulas for homotopy transfer. 
We view these results as partial progress towards clarifying the origin of the kinematic algebra of gauge theories. The main challenge, however, lies still ahead: the $C_\infty$ algebra of Yang-Mills theory constitutes only the first layer of the much richer, and much more complicated, off-shell kinematic algebra, the BV$_\infty^\B$ algebra identified by Reiterer in \cite{Reiterer:2019dys}.

In this respect, the first ingredient to go beyond the $C_\infty$ sector is a second nilpotent differential of degree $-1$. This operator plays the role of a BV Laplacian and defines the bracket of the kinematic algebra as its associated BV antibracket. In our worldline  setting, the homotopy BV Laplacian is given by the antighost $b$, which acts as a simple degree shift in the space of states $\cK$ carrying the $C_\infty$ algebra structure. Despite its simplicity, determining its \emph{order} with respect to $m_2$ is highly nontrivial, given that the latter is not a pointwise product. This has been a major  hurdle in the recent attempts to explicitly construct the kinematic BV$_\infty^\B$ algebra. 

To simplify this state of affairs, a possible route would be to find an analogous operator at the level of the quantized phase space, since it is generally simpler to determine the order of an operator with respect to $\star$. The antighost $b$ itself is a coordinate of the phase space, so a natural ansatz for such an operator would be the adjoint action $[b,-]_\star=\frac{\del}{\del c}$, which is a derivation of $\star$, i.e.~it is first order with respect to $\star$. However, $\frac{\del}{\del c}$ is \emph{not} an endomorphism of the subalgebra $\cF$, meaning that $[b,\cF]_\star\nsubseteq\cF$. Nevertheless, given the established homotopy equivalence with $\cK$, $b$ on $\cK$ can be transported to an endomorphism in $\cF$. 
To this end one first views 
$b$ as an operator $\hat b$ acting on $\cK$ and then defines 
\begin{equation}
 b_\cF:=I\circ\hat b\circ\Pi\in{\rm End}(\cF)\;.
\end{equation} 
It remains to be seen whether this operator, or perhaps a suitable modification of it, will help simplify  the BV$_\infty^\B$ algebra. 

More generally, part of the main challenge is 
to  guess what the proper algebraic structure is that in its ‘strict’ form captures all of color-kinematics duality.
For instance, one may speculate 
that the framework of factorization algebras may be useful, as there are very simple such algebras that are `quasi-isomorphic' to the kind of $\star$ product algebras that are at the heart of the present results  \cite{Chiaffrino:2024neo}. 
We plan to explore these and other directions in the future.

\section*{Acknowledgments}

We would like to thank Giuseppe Casale and Rob Klabbers for discussions and collaborations on closely related topics. The work of R.B. is funded by the Deutsche Forschungsgemeinschaft (DFG, German Research Foundation) – Projektnummer 524744955, “Worldline approach to the double copy”; the work of C.C. is funded by the Deutsche Forschungsgemeinschaft (DFG, German Research Foundation), ”Homological Quantum Field Theory”, Projektnummer 9710005691. M.F.K. is funded by the DFG – Projektnummer 417533893/GRK2575 “Rethinking Quantum Field Theory”.

\providecommand{\href}[2]{#2}\begingroup\raggedright\endgroup

\end{document}